\documentstyle{article}

\newcommand {\tsub}[1]{_{\mbox{\protect\scriptsize #1}}}
\newcommand {\tsup}[1]{^{\mbox{\protect\scriptsize #1}}}
\newcommand {\Ref}[1]{(\ref{#1})}
\newcommand 
{\Stretch}[1]{\renewcommand{\baselinestretch}{#1}\large\normalsize}

\newcommand {\ket}[1]{\left|#1\right\rangle}
\newcommand {\bra}[1]{\left\langle#1\right|}
\newcommand {\inner}[2]{ \left\langle #1 \right|
                         \left. #2 \right\rangle }

\begin{document}

\Stretch{3}
\begin{center}

{\large 17/7/96}
\vspace*{2em}

{\huge \bf Density Matrix Renormalisation Group for a Quantum Spin Chain 
at Non-Zero Temperature}
\vspace*{5em}

\Stretch{1.5}
{\large
R.\ J.\ Bursill$^{1\dagger}$,
T. Xiang$^{2}$ and G.\ A.\ Gehring$^{1}$
}
\vspace*{2em}

$^{1}$Department of Physics,
The University of Sheffield,
S3 7RH, Sheffield, United Kingdom.\\

$^{2}$Interdisciplinary Research Centre in Superconductivity,
The University of Cambridge, Cambridge, CB3 0HE, UK.\\

$\dagger$ Present address: School of Physics,
UNSW, Sydney, 2052, Australia.

\end{center}


\Stretch{1.2}

\begin{abstract}

We apply a recent adaptation of White's density matrix renormalisation 
group (DMRG) method to a simple quantum spin model, the dimerised $XY$ 
chain, in order to assess the applicabilty of the DMRG to quantum 
systems at non-zero temperature. We find that very reasonable results 
can be obtained for the thermodynamic functions down to low temperatures 
using a very small basis set. Low temperature results are found to be 
most accurate in the case when there is a substantial energy gap.

\end{abstract}

\setcounter{page}{0}


Since it's recent inception, White's denstiy matrix renormalisation 
group (DMRG) method has been established as the method of choice for 
determining static, low energy properties of one-dimensional quantum 
lattice systems \cite{development}--\cite{bursill}. Extensions
to the calculation of 
dynamical properaties \cite{dynamics} and even to the study of low 
temperature properties of two dimensional systems \cite{two_dimensions} 
have been forthcoming. Moreover, Nishino's formulation of the DMRG for 
two dimensional classical systems \cite{nishino} has paved the way for 
the study of one dimensional quantum systems at non-zero temperature. In 
this letter we present what is, to the best of our knowledge, the first 
application of the DMRG to the thermodynamics of a quantum system.

The system that we consider is a simple spin chain model---the 
dimerised, $S=1/2$, $XY$ 
model
\begin{equation}
{\cal H} = -\sum_{i=1}^{N/2}
\left[ 
J_{1} \left( S_{2i-1}^{x} S_{2i}^{x} + S_{2i-1}^{y} S_{2i}^{y} \right)
+
J_{2} \left( S_{2i}^{x} S_{2i+1}^{x} + S_{2i}^{y} S_{2i+1}^{y} \right)
\right]
\label{hamiltonian}
\end{equation}
where $S_{i}$ is a spin-1/2 operator for site $i$ on an (even) chain of 
$N$ sites, with periodic boundary conditions.

This model is exactly solvable \cite{exact_solution}; the Helmholtz free 
energy $\psi$ is given by
\begin{equation}
-\beta\psi = \lim_{N \rightarrow \infty} \frac{ \log Z_{N} }{ N }
= \frac{1}{2 \pi} \int_{0}^{2 \pi} \log \left[
2 \cosh \frac{\beta \phi(\theta)}{4}
\right] \, d \theta
\label{psi}
\end{equation}
where $Z_{N} = {\mbox Tr}\, e^{-\beta{\cal H}}$
is the partition function, $\beta \equiv 1/T$ is the inverse 
temperature,
$\phi(\theta) =
\cos \xi(\theta) + \gamma \cos ( \theta + \xi(\theta) )$,
$\xi(\theta) =
-\tan^{-1} \frac{\gamma \sin \theta}{1 + \gamma \cos \theta}$,
$\gamma \equiv J_{2}/J_{1}$ and we have set $J_{1} = 1$.
Thermodynamic properties such as the internal energy
$u = - \frac{\partial\,\beta\psi}{\partial\beta}$ and the specific heat
$c_{V} = -\beta^{2} \frac{\partial u}{\partial\beta}$ are readily 
obtainable from \Ref{psi}.

Because of its solvability and the fact that it possesses both gapless 
$(\gamma = 1)$ and gapped $(\gamma \neq 1)$ excitations, 
\Ref{hamiltonian} presents a useful test model for the extension of the 
DMRG to quantum systems at non-zero temperature.

Now Nishino's formulation of the DMRG applies directly to classical, two 
dimensional spin systems and so in order to apply it to the quantum 
system 
\Ref{hamiltonian}, we must first invoke the Trotter-Suzuki method 
\cite{suzuki}. That is, we make the decomposition ${\cal H} =
{\cal H}_{1} + {\cal H}_{2}$ where
\begin{equation}
{\cal H}_{1} = -J_{1} \sum_{i=1}^{N/2} h_{2i-1,2i},
\hspace*{2em}
{\cal H}_{2} = -J_{2} \sum_{i=1}^{N/2} h_{2i,2i+1}
\label{H1H2}
\end{equation}
and $h_{i,j} \equiv S_{i}^{x} S_{j}^{x} + S_{i}^{y} S_{j}^{y}$.
We then apply the formula
\begin{equation}
Z_{N} = \lim_{M \rightarrow \infty} Z_{MN}
\equiv
\lim_{M \rightarrow \infty}
{\mbox Tr}\,
\left[
e^{-\beta{\cal H}_{1}/M}
e^{-\beta{\cal H}_{2}/M}
\right]^{M}
\label{ZMN}
\end{equation}
Inserting a collection ${\sigma}$ of $2M$ complete sets of states into 
\Ref{ZMN} then yields
\begin{equation}
Z_{MN}
=
\sum_{{\sigma}}
\prod_{j=1}^{M}
\bra{ \sigma_{1}^{2j-1}\ldots\sigma_{N}^{2j-1} }
e^{-\beta{\cal H}_{1}/M}
\ket{ \sigma_{1}^{2j}\ldots\sigma_{N}^{2j} }
\bra{ \sigma_{1}^{2j}\ldots\sigma_{N}^{2j} }
e^{-\beta{\cal H}_{2}/M}
\ket{ \sigma_{1}^{2j+1}\ldots\sigma_{N}^{2j+1} }
\end{equation}
where periodic boundary conditions
$\sigma_{i}^{2M+1} \equiv \sigma_{i}^{1}$,
$\sigma_{N+1}^{j} \equiv \sigma_{1}^{j}$ are assumed.

Now, because the $N/2$ terms in the sums \Ref{H1H2} commute and act on 
different pairs of sites, we have
\begin{equation}
Z_{MN}
=
\sum_{{\sigma}}
\prod_{i=1}^{N/2}
\prod_{j=1}^{M}
\tau_{1}( \sigma_{2i-1}^{2j-1} \sigma_{2i}^{2j-1} | \sigma_{2i-1}^{2j} 
\sigma_{2i}^{2j} )
\tau_{2}( \sigma_{2i}^{2j} \sigma_{2i+1}^{2j} | \sigma_{2i}^{2j+1} 
\sigma_{2i+1}^{2j+1} )
\end{equation}
where
\begin{equation}
\tau_{i}(\sigma'\mu'|\sigma\mu)
\equiv
\bra{\sigma'\mu'} e^{\beta J_{i}h_{1,2}/M} \ket{\sigma\mu}
\label{taui}
\end{equation}
for $i=1,2$. Moreover,
\begin{eqnarray}
Z_{MN}
& = &
\sum_{{\sigma}}
\prod_{i=1}^{N/2}
{\cal T}_{1}( \sigma_{2i-1}^{1}\ldots\sigma_{2i-1}^{2M} | 
\sigma_{2i}^{1}\ldots\sigma_{2i}^{2M} )
{\cal T}_{2}( \sigma_{2i}^{1}\ldots\sigma_{2i}^{2M} |
\sigma_{2i+1}^{1}\ldots\sigma_{2i+1}^{2M} )
\\
& = &
\mbox{Tr}\,{\cal T}^{N/2}
\end{eqnarray}
where ${\cal T} \equiv {\cal T}_{1}{\cal T}_{2}$ is the {\em virtual 
transfer matrix} and
\begin{eqnarray}
{\cal T}_{1}
(\mu^{1}\ldots\mu^{2M} | \sigma^{1}\ldots\sigma^{2M})
& \equiv &
\prod_{j=1}^{M}
\tau_{1}( \mu^{2j-1}\sigma^{2j-1} |
\mu^{2j}\sigma^{2j} )
\\
{\cal T}_{2}
(\mu^{1}\ldots\mu^{2M} | \sigma^{1}\ldots\sigma^{2M})
& \equiv &
\prod_{j=1}^{M}
\tau_{2}( \mu^{2j}\sigma^{2j} |
\mu^{2j+1}\sigma^{2j+1} )
\end{eqnarray}
The matrices ${\cal T}_{1}$ and ${\cal T}_{2}$ are depicted graphically 
in Fig.\ 1.

It follows \cite{suzuki} that
\begin{equation}
\psi =  \psi^{(M)}
\equiv
\lim_{M\rightarrow\infty}
-\frac{ \log\lambda\tsub{max} }{ 2\beta }
\label{psiM}
\end{equation}
where $\lambda\tsub{max}$ is the eigenvalue of ${\cal T}$ with maximal 
modulus and generally depends on the Trotter dimension $M$.

We may now apply the formulation of the DMRG for transfer matrices to 
the calculation of $\lambda\tsub{max}$. We commence by defining an 
initial transfer matrix ${\cal T}\tsub{s}$ for a single-site {\em system 
block}, which connects it to adjacent sites to the left and the right 
viz
\begin{equation}
{\cal T}\tsub{s}( \sigma' n' \mu' | \sigma\, n\, \mu ; \sigma'' \mu'' )
=
\sum_{n''}
\tau_{1}( \sigma' \sigma'' | n' n'' )
\tau_{2}( n'' n | \mu'' \mu )
\end{equation}
The initial block, having just one site, has $m=2$ states, $n=1$ or 2 
($\uparrow$ or $\downarrow$) (see Fig.\ 2).

We next define an initial {\em environment block} and associated 
transfer matrix ${\cal T}\tsub{e}$ in precisely the same way:
${\cal T}\tsub{e} \equiv {\cal T}\tsub{s}$. We define a {\em superblock} 
using the system and environment blocks in addition to two added sites, 
arranged in a periodic fashion (see Fig.\ 3). That is, the superblock 
transfer matrix is given by
\begin{eqnarray}
{\cal T} ( \sigma_{1}' n_{1}' \sigma_{2}' n_{2}' |
\sigma_{1} n_{1} \sigma_{2} n_{2} )
& = &
( {\cal T}_{1} {\cal T}_{2} )
( \sigma_{1}' n_{1}' \sigma_{2}' n_{2}' |
\sigma_{1} n_{1} \sigma_{2} n_{2} )
\\
& = &
\sum_{\sigma_{1}'' \sigma_{2}''}
{\cal T}\tsub{s}
( \sigma_{1}' n_{1}' \sigma_{2}' | \sigma_{1} n_{1} \sigma_{2} ;
\sigma_{1}'' \sigma_{2}'' )
{\cal T}\tsub{e}
( \sigma_{2}' n_{2}' \sigma_{1}' | \sigma_{2} n_{2} \sigma_{1} ;
\sigma_{2}'' \sigma_{1}'' )
\label{superblock_transfer}
\end{eqnarray}
At this point ${\cal T}$ may be diagonalised to determine 
$\lambda\tsub{max}$ and hence the $M=2$ approximation \Ref{psiM} to 
\Ref{psi}.

In order to proceed to larger lattices, we must augment (expand) the 
system and environment blocks. We let $n \longleftrightarrow (p,\nu)$, 
$p=1,\ldots,m$, $\nu=\uparrow,\downarrow$ denote a state for an 
augmented system block, consisting of the initial system block and one 
site added to the right. There are now $m' = 2m = 4$ states:
$n = 1 \longleftrightarrow (1,\uparrow)$,$\ldots$,
$n = 4 \longleftrightarrow (2,\downarrow)$.

The transfer matrix ${\cal T}\tsub{s}'$ for the augmented system block 
is defined (see Fig.\ 4) by
\begin{equation}
{\cal T}\tsub{s}'
( \sigma' n' \mu' | \sigma\, n\, \mu ; \sigma'' \mu'' )
\equiv
\sum_{\nu''}
{\cal T}\tsub{s}
( \sigma' p' \nu' | \sigma\, p\, \nu ;
\sigma'' \nu'' )
\tau_{1} ( \nu' \nu'' | \mu' \mu'' )
\end{equation}
An augmented environment block is defined in a similar way, this time 
adding a site to the left viz
\begin{equation}
{\cal T}\tsub{e}'
( \sigma' n' \mu' | \sigma\, n\, \mu ; \sigma'' \mu'' )
\equiv
\sum_{\nu''}
{\cal T}\tsub{e}
( \sigma' p' \nu' | \sigma\, p\, \nu ;
\sigma'' \nu'' )
\tau_{2} ( \sigma'' \sigma | \nu'' \nu )
\end{equation}
Now, the superblock and its associated transfer matrix 
\Ref{superblock_transfer} can once again be formed using the augmented 
blocks with $m' \longmapsto m$,
${\cal T}\tsub{s}' \longmapsto {\cal T}\tsub{s}$ and
${\cal T}\tsub{e}' \longmapsto {\cal T}\tsub{e}$. Moreover, the process 
can be iterated, each time augmenting the system and environment blocks 
to the right and the left respectively.

However, in order to prevent the superblock basis from becoming too 
large, we must truncate it by capping the number of system (and 
environment) block states, $m$, which in principle doubles every 
iteration. To do so, we form reduced density matrices for the augmented 
blocks by performing an appropriate partial trace on the projection 
operator $\ket{\psi\tsub{max}}\bra{\psi\tsub{max}}$ formed from the 
eigenstate $\ket{\psi\tsub{max}}$ of ${\cal T}$ corresponding to the 
eigenvalue $\lambda\tsub{max}$.

That is, for the augmented system block, the density matrix 
$\rho\tsub{s}$ is defined by
\begin{equation}
\rho\tsub{s}(n'|n)
\equiv
\sum_{\sigma_{1} n_{2}}
\inner{\sigma_{1} n_{1}' \sigma_{2}' n_{2}}{\psi\tsub{max}}
\inner{\psi\tsub{max}}{\sigma_{1} n_{1} \sigma_{2} n_{2}}
\end{equation}
where $n \longleftrightarrow (n_{1},\sigma_{2})$ and
$n' \longleftrightarrow (n_{1}',\sigma_{2}')$. For the augmented 
environment block we have
\begin{equation}
\rho\tsub{e}(n'|n)
\equiv
\sum_{\sigma_{1} n_{1}}
\inner{\sigma_{1} n_{1} \sigma_{2}' n_{2}'}{\psi\tsub{max}}
\inner{\psi\tsub{max}}{\sigma_{1} n_{1} \sigma_{2} n_{2}}
\end{equation}
with $n \longleftrightarrow (n_{2},\sigma_{2})$ and
$n' \longleftrightarrow (n_{2}',\sigma_{2}')$.
Eigenvalues and eigenvectors of the density matrices are then found viz
\[
\{
\omega_{n}\tsup{(s)}, \ket{n}\tsup{(s)}:\;n=1,\ldots,m'
\}
\mbox{ and }
\{
\omega_{n}\tsup{(e)}, \ket{n}\tsup{(e)}:\;n=1,\ldots,m'
\}
\]
for $\rho\tsub{s}$ and $\rho\tsub{e}$ respectively where
$1 \geq \omega_{1}\tsup{(i)} \geq \omega_{2}\tsup{(i)}
\geq \ldots \geq \omega_{m'}\tsup{(i)}$ for
$\mbox{i}=\mbox{s}$
 or e.

In proceeding to the next iteration then, we represent the augmented 
block transfer matrices ${\cal T}\tsub{s}'$ and ${\cal T}\tsub{e}'$ in 
terms of the density matrix eigenvectors and truncate the block Hilbert 
spaces so as to retain only the $m$ most important states viz
\begin{equation}
\sum_{n'',n''' = 1}^{m'}
\inner{n'}{n''}\tsup{(i)}
{\cal T}\tsub{i}'
(\sigma' n'' \mu' | \sigma\, n''' \mu ; \sigma'' \mu'')
\;\tsup{(i)}
\inner{n'''}{n}
\longmapsto
{\cal T}\tsub{i}
(\sigma' n' \mu' | \sigma\, n\, \mu ; \sigma'' \mu'')
\end{equation}
for
$\mbox{i}=\mbox{s}$
and e, $n,n'=1,\ldots,m$ and where $m$ is the chosen cap on 
the number of states per block.

Now, ${\cal T}$ is a large, sparse, non-symmetric matrix, for which 
efficient algorithms for the determination of $\lambda\tsub{max}$ and 
$\ket{\psi\tsub{max}}$ are becoming available. So far we have simply 
used 
Schur decomposition and the inverse power method to perform this task. 
We have thus only been able to work with $m=16$. 
Results for $\psi$, $u$ and $c_{V}$ for $\gamma = 1$ (pure $XY$) and 
$\gamma = 2$ (dimerised $XY$) are given in
Figs 5--7.

Finite-$M$ results $\psi^{(M)}$ are obtained by fixing the inverse 
temperature $\beta=\beta_{0}$ in the expressions \Ref{taui} and 
increasing the superblock size $2M$, beginning with $M=2$, so at each 
iteration the temperature is identified as $T = 1/M\beta_{0}$. For the 
$\beta_{0}$ values we have considered (0.05--0.2) there is a small error 
due to the finiteness of the Trotter decomposition but this is 
negligible compared with the error due to Hilbert space truncation, a 
result which is easily verified by considering the exact solution for 
$\psi^{(M)}$ \cite{suzuki}. A typical calculation, using NAG library 
routines on a 333 MHz DEC Alpha machine, takes 6 hours to generate a 
superblock size of $2M = 300$.

We note that convergence at low temperatures is far better for the 
gapped system ($\gamma = 2$). However, even in the gapless case, the 
peak position and peak height in the specific heat are afflicted by 
errors of only ca.\ 3\%, and results may be improved markedly by using 
larger values of $m$ and the finite lattice method \cite{development}. 
Also, results for $u$ and $c_{V}$ are obtained by numerically 
differentiating spline fits of the computed $\psi^{(M)}$ values, and so 
improvements may be obtained by combining results from a number of 
different $\beta_{0}$ values.

Note that, at any given iteration, the DMRG usually works with a chain 
which is spatially finite. Here the chain is infinite and the finiteness 
is in the level of the Trotter approximation. Another difference is that 
the DMRG usually produces its best results for the ground state energy 
and less accurate results for higher excitations. A different situation
occurs here---the lower the temperature, the less accurate the result. 
Even so, as can be seen in Fig.\ 5, the exact ground state energy 
$\lim_{T\rightarrow 0}\psi$, is recovered with reasonable accuracy, 
especially in the gapped case $\gamma=2$.

Finally, we note that, as ${\cal T}$ is non-symmetric, the use of a 
truncated basis will not in general lead to a variational lower bound on 
$\lambda\tsub{max}$. However, in practice it can be seen, for instance, 
that the DMRG result for the internal energy $u$ is always bounded above 
the exact value.

To conclude, we have applied the DMRG to a quantum spin chain at
non-zero temperature, making use of Nishimo's adaptation of the method to 2D 
classical systems. Reasonable results are obtained for the specific heat 
down to low temperatures in calculations involving an extremely small 
basis set, agreement with the exact solution being markedly better in 
the case where the system has a substantial gap. The approach may prove 
useful in determining thermodynamic properties of models of 
experimentally realisable systems such as coupled chains and models with 
anisotropy, dimerisation and frustration.

R.\ J.\ B.\ gratefully acknowledges the support of SERC grant no.\ 
GR/J26748. Computations were performed on the DEC 8400
facility at CLRC.



\section*{Figure Captions}

\noindent
1. Pictorial representation of matrices a)
${\cal T}_{1}
(\mu^{1}\ldots\mu^{2M} | \sigma^{1}\ldots\sigma^{2M})$ and b)
${\cal T}_{2}
(\mu^{1}\ldots\mu^{2M} | \sigma^{1}\ldots\sigma^{2M})$.

\noindent
2. Pictorial representation of the initial system block transfer 
matrix
${\cal T}\tsub{s}
( \sigma' n' \mu' | \sigma\, n\, \mu ; \sigma'' \mu'' )$.
$n$ and $n'=1,\ldots,m=2$ are initial and final states of the
system block which consists of a single site.
$\sigma$, $\sigma''\mbox{ and }\sigma'=\;\uparrow\mbox{ or }\downarrow$ and
$\mu$, $\mu''\mbox{ and }\mu'=\;\uparrow\mbox{ or }\downarrow$ are initial,
intermediate and final states for the adjeacent sites to the
left and right of the block respectively. The intermediate
state $n''$ of the system block is summed over to produce
the matrix product.

\noindent
3. Pictorial representation of the superblock, which consists of 
system and environment blocks connected by two added sites to form a 
periodic chain. $n_{1}\mbox{ and } n_{1}'$ and $n_{2}\mbox{ and } 
n_{2}'$ are initial and final state indices for the system and 
environment blocks respectively.
$\sigma_{1}\mbox{ and }\sigma_{1}'$ and
$\sigma_{2}\mbox{ and }\sigma_{2}'$
are initial and final states for the two added sites. The intermediate 
states $\sigma_{1}''$ and $\sigma_{2}''$ are summed over to form the 
matrix product.

\noindent
4. Pictorial representation of the augmented (a) system and (b) 
environment blocks, which consist of the old blocks augmented by adding 
sites to the right and the left respectively. $n\mbox{ and } 
n'=1,\ldots,m'$ are the initial and final states of the new (augmented) 
block which consists of the old block ($p$ and $p'$) augmented with an 
added site ($\nu$ and $\nu'$). $\sigma$, $\sigma''\mbox{ and }\sigma'$ and 
$\mu$, $\mu''\mbox{ and }\mu'$ are state indices for the adjacent sites. 
The intermediate state $\nu''$ of the added site is summed over to 
produce the matrix product.

\noindent
5. Exact and DMRG results for the Helmholtz free energy $\psi$ as 
a function of temperature $T$ for two cases: $\gamma = 1$ and
$\gamma = 2$.

\noindent
6. Exact and DMRG results for the internal energy $u$ as a 
function of temperature $T$ for two cases: $\gamma = 1$ and
$\gamma = 2$.

\noindent
7. Exact and DMRG results for the specific heat $c_{V}$ as a 
function of temperature $T$ for two cases: $\gamma = 1$ and
$\gamma = 2$.

%
%
%
%
%
%
%

\begin{thebibliography}{99}

\bibitem{development}
S.\ R.\ White and R.\ M.\ Noack, Phys.\ Rev.\ Lett.\ 68 (1992) 3487;
S.\ R.\ White, Phys.\ Rev.\ Lett.\ 69 (1992) 2863,
Phys.\ Rev.\ B 48 (1993) 10 345.

\bibitem{bursill}
S.\ R.\ White and D.\ A.\ Huse, Phys.\ Rev.\ B 48 (1993) 3844;
E.\ S.\ S\o rensen and I.\ Affleck, Phys.\ Rev.\ Lett.\ 71 (1993) 1633;
R.\ J.\ Bursill, T.\ Xiang and G.\ A.\ Gehring,  J.\ Phys.\ A 28 (1994) 
2109;
R.\ J.\ Bursill, G.\ A.\ Gehring, D.\ J.\ J.\ Farnell, J.\ B.\ 
Parkinson, Tao Xiang and Chen Zeng, J.\ Phys.\ C 7 (1995) 8605;
S.\ R.\ White, R.\ M.\ Noack and D.\ J.\ Scalapino, 
Phys.\ Rev.\ Lett.\ 73 (1994) 886;
R.\ M.\ Noack, S.\ R.\ White and D.\ J.\ Scalapino,
Phys.\ Rev.\ Lett.\ 73 (1994) 882;
L.\ G.\ Caron and S.\ Moukouri, Phys.\ Rev.\ Lett.\ 76 (1996) 4050;
G.\ A.\ Gehring, R.\ J.\ Bursill and T.\ Xiang, {\em Applications of the 
density matrix renormalisation group to problems in magnetism}, To 
appear in Acta Physica Polonica (cond-mat/9608127).

\bibitem{dynamics}
H.\ B.\ Pang, H.\ Akhlaghpour and M.\ Jarrell, Phys.\ Rev.\ B 53 (1996) 
5086;
K.\ A.\ Hallberg, Phys.\ Rev.\ B 52 (1995) R9827.

\bibitem{two_dimensions}
S.\ Liang and H.\ Pang, Europhys.\ Lett.\ 32 (1995) 173;
T.\ Xiang, Phys.\ Rev.\ B 53 (1996) 10445;
S.\ R.\ White, {\em Spin Gaps in a Frustrated Heisenberg model for 
CaV$_4$O$_9$}, Preprint;
S.\ R.\ White and D.\ J.\ Scalapino, {\em Hole and Pair Structures in 
the $t$-$J$ model}, Preprint.

\bibitem{nishino}
T.\ Nishino, J.\ Phys.\ Soc.\ Jap.\ 64 (1995) 3598;
T.\ Nishino, K.\ Okunishi and M.\ Kikuchi, Phys.\ Lett.\ A, 213 (1996) 
69.

\bibitem{exact_solution}
L.\ N.\ Bulaevskii, Sov.\ Phys.\ JETP 17 (1963) 1008.

\bibitem{suzuki}
M.\ Suzuki and M.\ Inoue, Prog.\ Theor.\ Phys.\ 78 (1987) 787;
M.\ Inoue and M.\ Suzuki, Prog.\ Theor.\ Phys.\ 79 (1988) 645.

\end{thebibliography}
\end{document}